\documentclass[nonacm, sigconf]{acmart}

\usepackage{enumitem}
\usepackage{array} 
\usepackage{multirow}
\usepackage{lipsum}
\usepackage{wasysym}
\usepackage{makecell}
\usepackage[T1]{fontenc}
\usepackage[utf8]{inputenc}
\usepackage{xspace}
\newcolumntype{L}[1]{>{\raggedright\let\newline\\\arraybackslash\hspace{0pt}}m{#1}}
\newcolumntype{C}[1]{>{\centering\let\newline\\\arraybackslash\hspace{0pt}}m{#1}}
\newcolumntype{R}[1]{>{\raggedleft\let\newline\\\arraybackslash\hspace{0pt}}m{#1}}
\newcolumntype{J}[1]{>{\let\newline\\\arraybackslash\hspace{0pt}}m{#1}}

\usepackage[colorinlistoftodos,prependcaption,textsize=tiny]{todonotes}

\AtBeginDocument{%
  \providecommand\BibTeX{{%
    \normalfont B\kern-0.5em{\scshape i\kern-0.25em b}\kern-0.8em\TeX}}}

\begin{document}


\title{WebSPL: A Software Product Line for Web Applications}

\author{Maicon Azevedo da Luz}
\affiliation{%
  \institution{Universidade do Vale do Rio dos Sinos}
  \city{S\~ao Leopoldo}
  \state{RS}
  \country{Brazil}
}
\email{pnpinformatica@gmail.com}

\author{Kleinner Farias}
\affiliation{%
  \institution{Universidade do Vale do Rio dos Sinos}
  \city{S\~ao Leopoldo}
  \state{RS}
  \country{Brazil}
}
\email{kleinnerfarias@unisinos.br}

\renewcommand{\shortauthors}{Maicon Luz and Kleinner Farias}

\begin{abstract}
Companies developing Web applications have faced an increasing demand for high-quality products with low cost and production time ever smaller. However, developing such applications is still considered a time-consuming and error-prone task, mainly due to the difficulty of promoting the reuse of features (or functionalities) and modules, and the heterogeneity of Web frameworks. Nowadays, companies must face ever-changing requirements. Software product lines emerged as an alternative to face this challenge by creating a collection of applications from a core of software assets. Despite the potential, the current literature lacks works that propose a product line for Web applications. This paper, therefore, presents WebSPL, a product line for Web applications that supports the main features found in Wed applications in real-world settings. The proposed WebSPL was evaluated by comparing it with a Web application developed based on a traditional approach. A case study that involves the development of two Web applications enabled data collection. Two Web applications were developed --- one with and another without the support of the proposed WebSPL. We compared these two applications using software design metrics, including complexity, size, duplicate lines, and technical debt. The initial results were encouraging and showed the potential for using WebSPL to support the development of Web applications.
\end{abstract}



\maketitle

\section{Introduction}

Nowadays, companies have faced the challenge of developing Web applications with high quality, at low cost, and in a short time~\cite{oliveira2018brcode,carbonera2020software}. Such companies have as primary goals the constant search to increase the quality of the developed Web applications, act in different business domains, and be agile in production and delivery. Moreover, they seek to promote the reduction of cost development and increase productivity and use technologies that facilitate the customization of Web applications, avoiding highly coupling applications~\cite{urdangarin2021mon4aware}. Achieving these goals is decisive for such companies, as it will have a direct impact on their profits and, consequently, on their survival.

In this context, companies seek to adopt methodologies that enhance the systematic reuse of software product modules, allowing the construction of interchangeable modules between applications and favoring the lifecycle management of their products to be made available to customers \cite{Kang2010}. By adopting such practices, it is expected that the challenge of building high-quality products in a shorter period and at a lower cost will be overcome or, at least, minimized. Moreover, developing software has become a compositional practice~\cite{farias2010empirical, farias2013analyzing, farias2010assessing}, where developers produce artifacts by accommodating building blocks. Previous studies demonstrate this concern when reporting empirical studies on the integration of software artifacts~\cite{bischoff2019integration} and their stability~\cite{farias2014effects,farias2011evaluating}.

Currently, to face this problem, companies have been using traditional development methodologies, including development based on components, and services and based on the object-oriented paradigm. Although these methodologies have been widely used, when compared to Software Product Lines (SPL), they are not as competitive, or even productive, for the development of high-quality software at low cost and in a short period, especially when applied without traceability between the software artifacts created. It has been observed that such methodologies are strictly related to implementation issues such as, for example, the definition of subroutines, modules, objects, components, and services, as opposed to promoting a consistent alignment between the companies\rq business objectives and the software products offered by them. This misalignment has been critical for companies that develop and market Web applications, including business management Web systems and social networks.

Academia and industry have invested efforts to understand the features of such Web applications to systematize the development process. Previous studies \cite{Pohl2005} show evidence that companies produce and maintain product families that share common characteristics and that have some particularities that differentiate them. Given the inefficiency of traditional methodologies, software product lines (SPL)~\cite{pohl2005software} emerged as an alternative to face the challenge of creating a collection of applications from a core of software assets. SPL emerged as an approach considered promising for Web development, as it allows the alignment between the companies' business strategies and the software development paradigm, allowing for a gain in productivity by leveraging the traceability and, consequently, the reuse of software artifacts. To sum up, SPL is a promising approach for reusing knowledge and artifacts among similar software products~\cite{reinhartz2020extracting}.

Northrop~\cite{Northrop2008} shows that the use of SPL was decisive for the development of software products in large companies (such as Nokia and Microsoft), in the areas of command and control systems for terrestrial spacecraft, control systems and warship (known as Celsiustech), and Web systems for the stock market. The use of SPL has allowed us to systematically deal with: (1) different types of keys and screen sizes; (2) localization features applied in different regions around the world; (3) the internationalization of cell phones through support for multiple languages; (4) interoperability of cell phones through the support of multiple network protocols; (5) maintainability of compatibility between different versions of the company's products; (6) support for configurable device features; and (7) the need to manage product functionality during the development phase and their maintenance and evolution phase. In \cite{Kang2010}, the authors demonstrate that SPL enables the development of products that share the same architecture and, consequently, the components that configure this architecture; the difference between them being represented by their variability.

Although applications are increasingly developed and made available on the Web, such as e-commerce and government applications, their development is still costly and often exceeds established deadlines. That said, the need for studies that try to solve, or even mitigate, the challenge of developing high-quality Web applications with short deadlines and reduced budget is characterized. Therefore, the great contribution of this work is to use SPL to develop families of Web applications based on the traceability and systematic reuse of their modules, managing to produce new products faster through the effective management of their variant parts.

To overcome the challenge of developing high-quality web applications in less time and at a lower cost, it is necessary to solve three core problems. Given that Web applications usually share a common core, and these are differentiated by some particularities, developers need to adequately understand what the application's features are and how the variability between Web applications manifests itself. However, in practice, developers do not understand the features, as well as the variability between application products. This can be explained by some reasons: (1) the inadequate representation of application features and dependencies between them; (2) not specifying the differences between the products, as well as their variability; and (3) the lack of definition of which features are needed to configure a particular product. Therefore, the comprehension of application features is hampered by the inadequate representation of the features that make up the products, as well as their variability.

The lack of understanding can be explained by some reasons: (1) the features that make up the product family are rarely defined, specified, and validated, as well as mapped precisely to software artifacts. This implies that developers commonly do not have access to mapping features to structural and behavioral diagrams of applications. That is, developers find it difficult to navigate from artifacts that specify requirements, to artifacts that define the design and implementation of features, and vice versa. Therefore, there is an imprecision in the traceability between features, their artifacts, and the modules that implement them. The current literature fails to provide details and evidence that proves whether the use of SPL can bring real benefits, including greater reusability and better modularity of the generated products when compared to Web applications developed following traditional practices. It could be, for example, that the nature of the Web applications domain compromises the real expected gains with SPL.

 Despite the potential of software product lines, the current literature lacks works that propose a product line for Web applications. This paper, therefore, presents WebSPL, a product line for Web applications that supports the main features found in Wed applications in real-world settings. The proposed WebSPL was evaluated by comparing it with a Web application developed based on a traditional approach. A case study that involves the development of two Web applications enabled data collection. Two Web applications were developed (one with and another without the support of the proposed WebSPL). We compared these two applications using software design metrics, including complexity, size, duplicate lines, and technical debt. The initial results were encouraging and showed the potential for using WebSPL to support the development of Web applications.

The paper is organized as follows. Section~\ref{background} introduces the main concepts used in this work. Section~\ref{section:trabalhos_relacionados} presents a comparative analysis of related works. Section~\ref{section:linha_de_produto_proposta} introduces the proposed software product line. Section~\ref{section:avaliacao} discusses evaluation. Finally, Section~\ref{section:conclusao} describes some conclusions and future work.

\section{Background}
\label{background}

This section presents the main concepts needed to understand the proposed work.

\subsection{Software product lines (SPL)}
\label{subsection:Linha_de_Produto_de_Software}

\textbf{Software product lines}. SPL~\cite{pohl2005software} represents a software development methodology that seeks to define techniques to increase the productivity of organizations through the systematic reuse of software assets. This methodology aims to replace the \textit{ad hoc} reuse practiced until then with a systematic reuse approach, one that enhances and manages the reuse of organizational assets. Clements and Northrop~\cite{Clements2001} define SPL as a set of software systems that share a common and managed set of features, which satisfy specific needs of a market segment, and which are developed in a predefined way from a common set of assets. Czarnecki and Eisenecker~\cite{Czarnecki2000} define a feature as being ``a property of a system that is relevant to a customer and that is used to capture the common and different aspects between products in a line.'' Typically, these features are classified as \textit{mandatory}, \textit{optional}, and \textit{alternative}. Furthermore, systematic reuse is only achieved when it is possible to understand how a family of software systems share features; and understand how the differences between such systems manifest themselves.

The SPL concept offers a new strategy to allow advancement in the productivity gains of development teams in companies. This strategy tends to enhance reuse by identifying the common and different parts between the products in such a way that each product can be elaborated through the systematic reuse of the common/different parts between the products. By implementing an SPL~\cite{Pohl2005}, companies can (1) increase the productivity and quality of software products; (2) reduce development cost in a short time; (3) decrease product rework and delivery time; and (4) enhance companies\rq adaptation to reach new markets by being able to develop new products faster and with quality.

\textbf{Feature-oriented domain analysis.} Software development using SPL methodology can be divided into two steps: (1) domain engineering, where SPL requirements are defined, specified, and validated. The definition of the structure used in the product line occurs in domain engineering, defining the points that will be common to the products as well as the points of variability; and (2) application engineering, where the products in the line are analyzed, designed, and implemented.

The concept of Feature-Oriented Domain Analysis (FODA) emerged in 1990~\cite{Kang1990}. Feature-oriented modeling seeks to identify and analyze the requirements (or characteristics) that are common and the variability in an SPL. Feature modeling is the activity of identifying visible external features in the product line and arranging those features in a model. Features are also a way to identify product line requirements, which are functional or non-functional requirements. The variability of these features can be divided into mandatory  (common to different products), optional, and alternative.

\subsection{Variability}
\label{subsection:variabilidade}

Variability is essential to the product line and is associated with the ability to adapt to changes, more precisely with planned changes. It is introduced during the product management sub-process when common features and variables are identified in the product line. Domain requirements detail the features defined in the product management, thus the features exert a strong influence not only on domain requirements but also on design, development, and testing. It is possible to use different levels of abstraction, with each new level of abstraction the previous level is refined and new features are added, resulting in a refinement. For this, it is necessary to introduce variability from the architecture so that the components can be compatible with different versions of the product line~\cite{Pohl2005}.

Within this context, there are points of variability, a point of variability represents an opportunity for variation within the SPL domain. Variation points are modeled to allow customization of applications using defined reuse, allowing customizations to adjust to meet the proposed needs.

\subsection{Web Applications}
\label{subsection:aplicacoes_web}

Web applications allow anyone with a browser to use them. With the increasing popularity of web applications, it is possible to access not only common computers but also mobile devices such as tablets and smartphones. More and more web applications are developed and made available. In this scenario, it is important to note that reuse plays an important role. SPL has become a key factor for the successful development of Web applications. However, it is possible to observe the lack of specific technologies and the lack of strategy applied to the Web application domain~\cite{Zhou2010}.

Web applications have proven to be an excellent area for applying SPL, as it makes development more agile. According to \cite{Balzerani2005}, web applications can be considered software products derived from a common infrastructure, where the core is the domain abstraction, for example, shopping cart, login, user registration, and retail system.

\subsection{Platform Architecture}
\label{subsection:arquitetura_da_plataforma}

One of the most important assets of a product line is the architecture, known as Product Line Architecture (PLA), also known as the platform. Through a common platform, products can be customized, where it is possible to produce different products based on a platform. The architecture must make explicit the commonalities and variability, allowing the sharing of common SPL elements, and facilitating the customization of the products, it defines the main SPL assets that were identified during the domain analysis: requirements, design, implementation, testing, etc. It is a strategic point for the organization that seeks to adopt a product line approach, the creation or removal of a platform exerts a strong influence on business success~\cite{Pohl2005}.

The construction of the platform requires an initial investment by the organization, as it will be necessary to make an initial effort, only after its completion the products will be developed. The removal or addition of new features on the platform affects all products that use it, so it is possible to maintain and evolve the platform, allowing the organization to adapt to market needs. Other requirements that must be met by the platform are security and performance, the platform must abstract these particularities for the products.

\subsection{Software Metrics}
\label{subsection:metricas_de_software}

Software metrics are the way to measure software. Used as a tool to analyze the software produced, providing information that can assist in its development. Software metrics can be divided into two groups: software product metrics and software process metrics. While software product metrics are measures extracted from source code or code design, for example, software process metrics are metrics extracted from the process used to develop the software.

The use of metrics is important to guide the organization and assist in decision-making. It is recommended that the extraction of metrics be carried out from the early stages of the project~\cite{Li2010}. The extraction of PLA metrics has two objectives: to assess the quality and to serve as a basis for analyzing the management and economic value of the product line. Variability impact analysis can determine the added value of a product line to the organization. The application of metrics in PLA provides efficient indicators to assess whether PLA is fulfilling its role within the product line of which it is part \cite{OliveiraJunior2010}.

\section{Related work}
\label{section:trabalhos_relacionados}

This section aims to present and compare other studies that proposed or used software product lines, or some reuse methods. Through this comparison, it is possible to identify the similarities and opportunities to be explored to expand the research. Five works that are similar to this research were selected.

\subsection{Analysis of the Related Work}

\textbf{Aziz \textit{et al.} \cite{A_Web-based_Software_Product_Line_Engineering_Framework}.} This study argues that using software product line engineering practices as the basis of a web application framework can make web application development cheaper, faster, and better in terms of quality. In this sense, the study proposes a framework to solve issues found in the development of product lines. The worker uses \textit{Abstract Behavioral Specification }(ABS) and various SPL engineering tools to manage implementation variability and generate the final product. A feasibility study shows that the framework can be used to develop a family of web applications in the Adaptive Information System for Charitable Organizations domain.

\textbf{Constantino \textit{et al.} \cite{Multiple_View_Interactive_Environment_to_Analyze_Software_Product_Line_Tools}.} The study argues that the adoption of SPL in the industry depends a lot on technological support, mainly on the tool usage factor. The article presents a visualization environment, called ViSPLatform, which aims to portray data related to experiments focused on SPL tools. ViSPLatform has been evaluated to analyze the extent to which the platform is effective in supporting the understanding of the characteristics of SPL tools. It is noteworthy that the study does not propose an SPL for the development of Web applications, but an environment for visualization. Preliminary results reported in the study show that ViSPLatform can somehow indicate strengths and opportunities for improvement in the analyzed SPL tools. The study highlights that the SPLOT tool has automatic analysis as its strength, but the interface still needs improvement.

\textbf{Horcas \textit{et al.} \cite{Software_Product_Line_Engineering_A_Practical_Experience}.} It performs an analysis of different tools that support the development and use of SPL, evaluating within the main features of SPL if they meet the promised requirements and the improvement points found.

\textbf{Mendonca \textit{et al.} \cite{S.P.L.O.T._Software_Product_Lines_Online_Tools}.} It presents a web tool that allows the analysis and configuration of an SPL. In the end, it provides a repository of generated models to foster further research.

\textbf{Yoshida and Iwane \cite{Knowledge-based_experimental_development_of_Web_application_systems}.} It introduces a tool to generate code to instantiate new SPL for web applications, demonstrating its use through a real web application.

\subsection{Comparative Analysis and Opportunities}
\label{subsection:comparative_analysis_and_oportunities}

For the comparison between the related works, the Comparison Criteria (CC) were established, the results are in Table~\ref{table:comparative_analysis_of_the_related_wprks_selected}.

\begin{itemize}
     \item \textbf{Web Applications (CC01):} Studies related to the development of web applications;
     \item \textbf{Case Study/Prototype (CC02):} Papers that used a case study or developed a prototype;
     \item \textbf{No code generation (CC03):} Searches that do not involve code generation to instantiate the SPL.
     \item \textbf{Assessment tools (CC04):} Studies that applied or created a form of SPL analysis involving software metrics;
\end{itemize}

\begin{table}[!ht]
    \footnotesize
    \caption{Comparative analysis of related works}
    \label{table:comparative_analysis_of_the_related_wprks_selected}
    \centering
    \begin{tabular}
    {
        J{0.35\linewidth}
        C{0.05\linewidth}
        C{0.05\linewidth}
        C{0.05\linewidth}
        C{0.05\linewidth}
    }
        \toprule
      \multicolumn{1}{c}{\multirow{2}[2]{*}{Related Work}} & \multicolumn{4}{c}{Comparison Criterion}
      \\\cline{2-5}
      & CC01 & CC02 & CC03 & CC04 \ \\

        \midrule
        Proposed work       & $\CIRCLE$    & $\CIRCLE$     & $\CIRCLE$          & $\CIRCLE$ \\
        Yoshida and Iwane \cite{Knowledge-based_experimental_development_of_Web_application_systems}                          & $\CIRCLE$    & $\CIRCLE$     & $\Circle$          & $\Circle$ \\
        
        Mendonca \textit{et al.} \cite{S.P.L.O.T._Software_Product_Lines_Online_Tools}         
        & $\LEFTcircle$    & $\Circle$     & $\Circle$          & $\Circle$ \\
        
        Horcas \textit{et al.} \cite{Software_Product_Line_Engineering_A_Practical_Experience}       
        & $\CIRCLE$    & $\Circle$      & $\Circle$         & $\Circle$ \\
        
        Constantino \textit{et al.} \cite{Multiple_View_Interactive_Environment_to_Analyze_Software_Product_Line_Tools}  
        & $\LEFTcircle$    & $\Circle$     & $\Circle$          & $\Circle$ \\
        
        Aziz \textit{et al.} \cite{A_Web-based_Software_Product_Line_Engineering_Framework}         
        & $\CIRCLE$    & $\CIRCLE$     & $\Circle$          & $\Circle$ \\
        
        \bottomrule
    \end{tabular}
        
    \footnotesize
    \begin{tabular}{ccc}
         $\CIRCLE$ Completely fills & $\LEFTcircle$ Partially fills & $\Circle$ Does not fill
    \end{tabular}
\end{table}

\textbf{Research opportunities.} Based on the comparative analysis we highlight some research opportunities: (1) no study proposes a software product line; (2) the main features of a Web Application are not revealed; (3) a case study supported by metrics was not performed, aiming to produce empirical knowledge and (4) the use of real-world technology to develop an SPL for Web applications is still scarce.

\section{Proposed Software Product Line}
\label{section:linha_de_produto_proposta}

This section introduces WebSPL, a product line of Web applications. WebSPL was designed and implemented following good practices documented in the software product line literature~\cite{pohl2005software, clements2002}. Thus, the design and implementation of WebSPL proposed in this article have two stages: domain engineering and application engineering. Section~\ref{subsection:engenharia_de_dominio} explains how WebSPL's domain engineering was performed, describing the features that make up WebSPL. Section~\ref{subsection:engenharia_de_aplicacao} describes the application engineering, highlighting how the proposed product line was implemented.

\subsection{Domain Engineering}
\label{subsection:engenharia_de_dominio}

The Domain Engineering stage focuses on eliciting requirements, specifying variability, and defining the configuration of the products in the line. Thus, a set of activities are performed, including (1) the description of the proposed SPL requirements, as well as the possible versions of the line; (2) the identification of SPL variability and specification of the feature diagram; and (3) description of product configurations that may be derived from SPL.

\textbf{Description of requirements and versions.} The development of the proposed product line uses a proactive approach, which according to \cite{Kang2010}, is suitable for mature and stable domains, allowing the characteristics of the product line to be planned. The construction of the product line was divided into versions, allowing the addition of new features and improving existing features as each new version is released. The use of versions is also focused on modularity, allowing stable and functional versions of the product line to be released at each new version, facilitating the maintenance and evolution of the features that are developed.

\begin{figure*}[]
    \centering
    \includegraphics[scale=0.23]{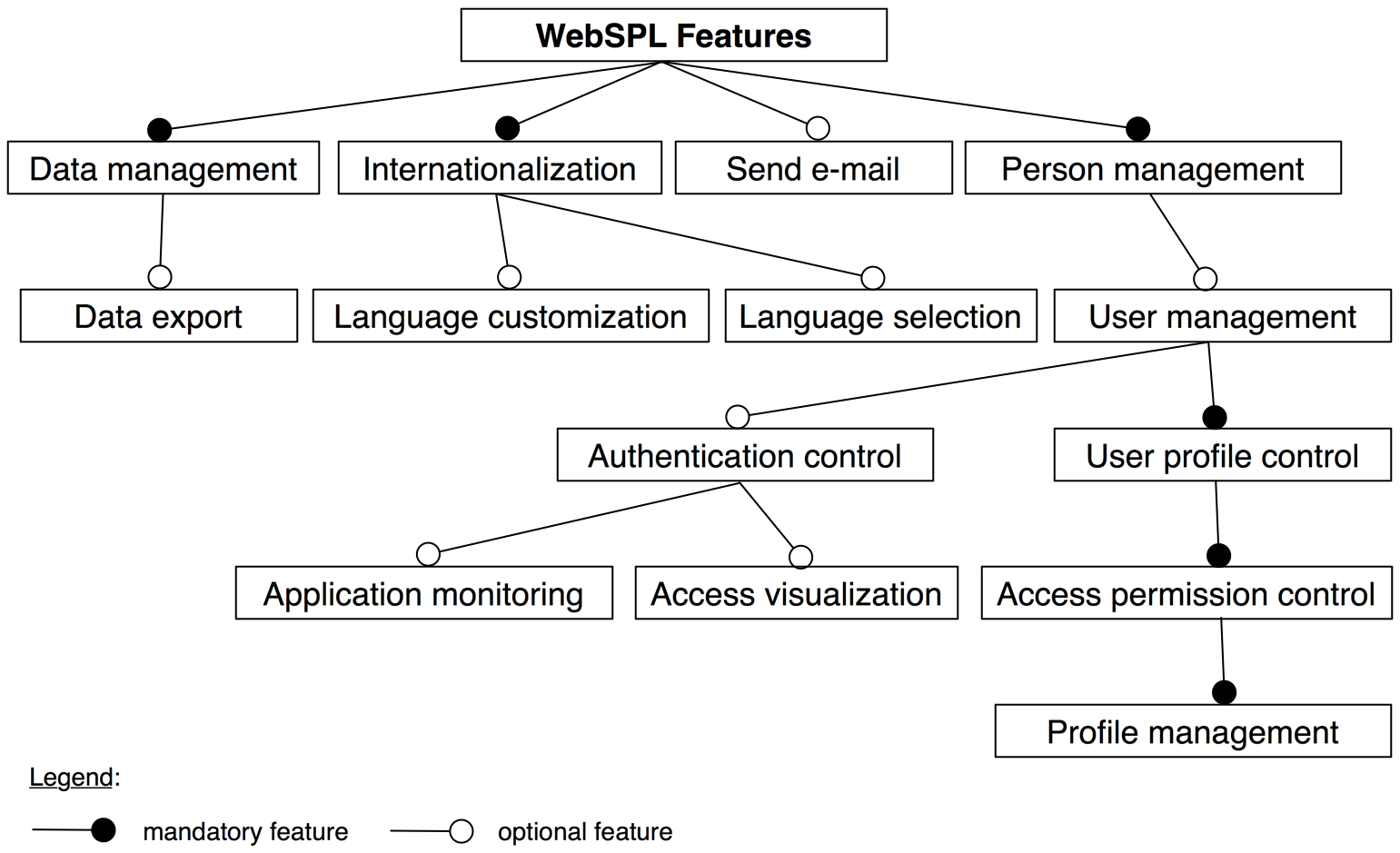}
    \caption{Feature diagram}
    \label{figure:diagrama_de_features}
\end{figure*}

The first version of the product line is based on some common features in a web application, the ability to manage and maintain the information handled by web applications is a basic feature, and serves as the basis for the other functionalities. The other features of this version focus on managing people and supporting internationalization. To facilitate the visualization of the versions, Table~\ref{table:versos_da_linha_de_produto_de_software} summarizes the features of each version and Figure \ref{figure:diagrama_de_features} presents the features diagram.

\begin{table}[!ht]
    \footnotesize
    \caption{Software Product Line Versions}
    \label{table:versos_da_linha_de_produto_de_software}
    \centering
    \begin{tabular}{l p{14.3pc}}
        \toprule
            Version & Description \\
        \midrule
Version 1 & Mandatory and Most Important Features for the Web Applications Product Line. \\
Version 2 & Addition of the remaining optional features related to user management. \\
Version 3 & Added mandatory features that relate to optional feature management
adds the ability to manage user permissions. \\
Version 4 & Added an optional feature for data export. Refactoring of other features to provide the export option. \\
        \bottomrule
    \end{tabular}
\end{table}

\begin{figure}[]
    \centering
    \includegraphics[width=0.92\columnwidth]{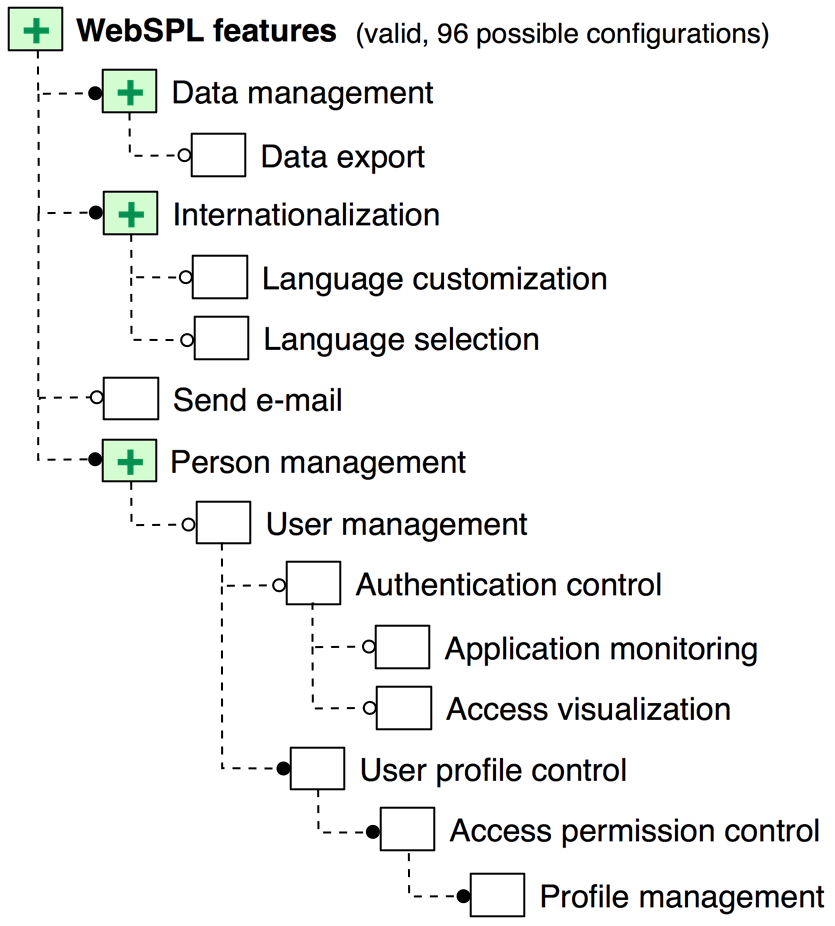}
    \caption{Configuration of a product with mandatory features} 
\label{figure:Configuracao_de_um_produto_com_as_features_obrigatorias}
\end{figure}

\begin{figure}[]
    \centering
    \includegraphics[width=0.92\columnwidth]{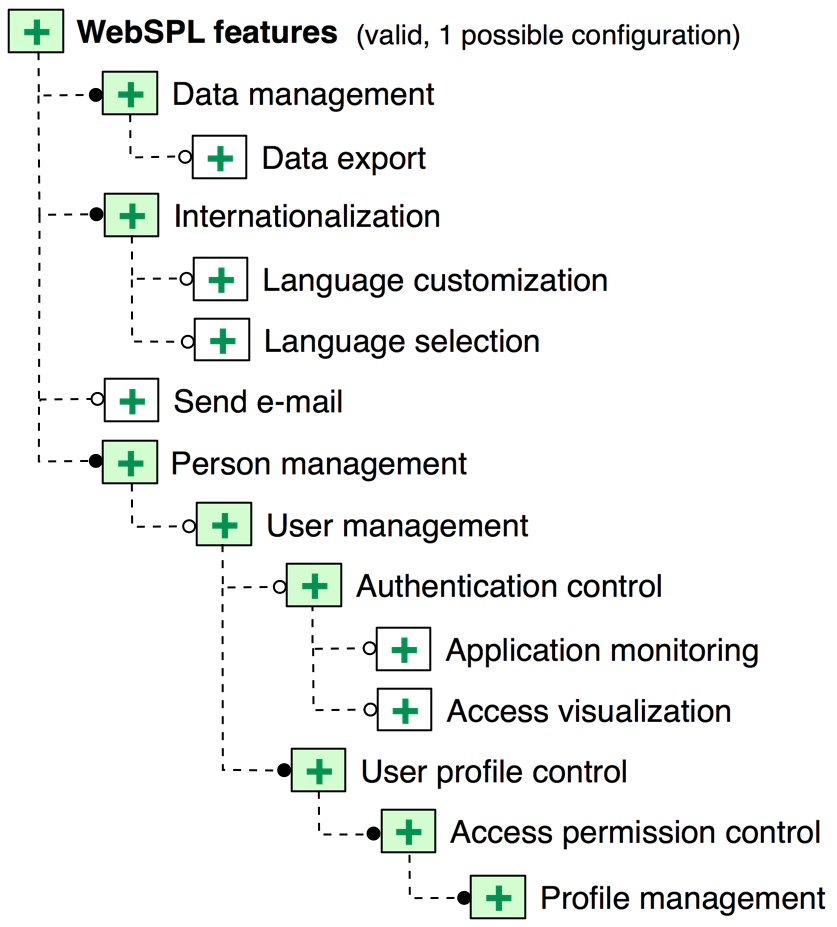}
    \caption{Configuration of a product with all the features}
    \label{figure:Configuracao_de_um_produto_com_todas_as_features}
\end{figure}

\textbf{Product configurations.} The product line configuration has as its main objective to define a product configuration, as well as allow the identification of possible product configurations. Figure~\ref{figure:diagrama_de_features} presents the WebSPL feature diagram. WebSPL has four built-in features, including data management, internationalization, user profile control, and profile management. Based on the feature diagram presented, two configurations of two products are presented. The product configurations were generated using the FODA notation (\textit{feature-oriented domain analysis}) through the FeatureIDE~\cite{featureIDE} tool. Figure \ref{figure:Configuracao_de_um_produto_com_as_features_obrigatorias} and Figure \ref{figure:Configuracao_de_um_produto_com_todas_as_features} show examples of possible variability with the features of the proposed SPL.

\subsection{Application Engineering}
\label{subsection:engenharia_de_aplicacao}

To carry out the implementation of WebSPL, the Java language was chosen. For the development of Web applications, the Java language allows the use of several frameworks, such as Spring Tools, JPA, Maven, and JSF, among others. Some of these frameworks are present in the language's standard API, while others can be incorporated into the application to increase productivity, add new functionality and implement an API specification.

\textbf{Context and dependency injection.} Seeking a loose coupling between the features and a greater facility to integrate them into the products, we chose to use dependency injection in all features. Fowler~\cite{Fowler2004} recommends when classes are used by multiple applications. One of the approaches used to control feature dependencies and variability is the use of aspect-oriented programming (AOP). However, it is known that AOP has some limitations, such as code with low comprehensibility, difficulty in maintaining the scope of performance of the pointcuts of the applications, and the overlapping of aspects. Another point that should be noted in the case of the Java language is the lack of a standard API for AOP in the language, forcing developers to choose a single framework, and losing code portability, as each framework implements AOP differently. Given this scenario, we chose to use Context and Dependency Injection (CDI), as it is a standard API in Java from Version 6 of Java EE. CDI provides the dependency injection mechanism as well as mechanisms that can replace the use of AOP as events and interceptors.

\begin{figure*}[]
    \centering
    \includegraphics[scale=0.2]{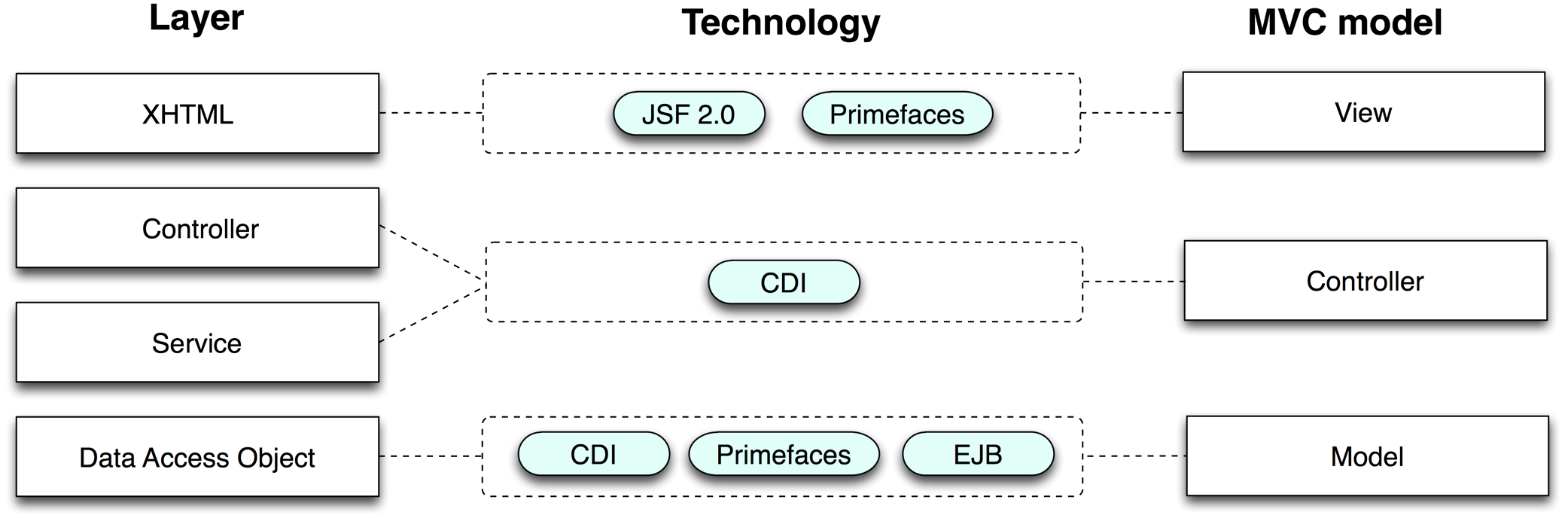}
    \caption{The architecture of the proposed SPL}
    \label{figure:arquitetura_da_LPS}
\end{figure*}

\textbf{WebSPL's proposed architecture}. Figure~\ref{figure:arquitetura_da_LPS} presents the WebSPL architecture, highlighting the layers, technologies used, and the MVC (model-view-controller) model used as a basis.

The presentation layer (view) of the features uses Java Server Faces\footnote{JSF: https://www.oracle.com/java/technologies/javaserverfaces.html} (JSF) together with the component library Primefaces\footnote{Primefaces: https://www.primefaces.org/}. This approach allows the final product to be customized with another look through the use of Cascading Style Sheets (CSS) if necessary. It was defined that SPL supports two languages by default, Brazilian Portuguese (pt\textunderscore BR) and American English (en\textunderscore US), but SPL allows the final product to be configured with the other language or have the messages already defined by default, allowing the user to choose the language. The central layer uses CDI to implement controllers and services. Information persistence uses Hibernate as a persistence framework, allowing SPL to run on different databases.

The WebSPL is divided into four layers (Figure \ref{figure:arquitetura_da_LPS}). Each layer has its responsibilities described as follows:

 \begin{itemize}
\item \textbf{EXtensible Hypertext Markup Language (XHTML)}: Contains the visual components of the SPL that will be displayed in the browser;
    
     \item \textbf{Controller}: It implements the user navigation control logic, transfers requests to the service layer, and interacts with the user by presenting success or failure messages;
    
     \item \textbf{Service}: It implements SPL services, business rules, and SPL operating logic;
    
     \item \textbf{Data Access Object (DAO)}: It implements the database and transaction access logic.
\end{itemize}

\begin{figure}[]
    \centering
    \includegraphics[scale=0.7]{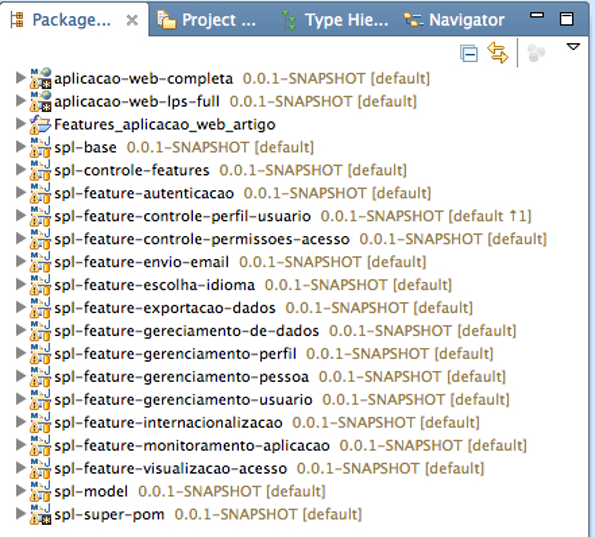}
    \caption{Projects that make up the WebSPL}
    \label{figure:projetos_que_compoem_a_LPS}
\end{figure}

\textbf{Dependency management.} Dependency management occurs through Maven \cite{Apache2014}, thus allowing dependencies between features considered mandatory to occur automatically. This is one of the main features of Maven and it proves to be suitable for the SPL environment, as it facilitates the configuration of the products. In this way, the SPL features were logically separated into independent Java projects (Figure \ref{figure:projetos_que_compoem_a_LPS}), allowing easy reading of the implemented features.

Another feature used by Maven is the integration with the Sonar quality inspection and analysis tool, performing an immediate analysis of code quality through software metrics.

\begin{figure}[]
    \centering
    \includegraphics[scale=0.16]{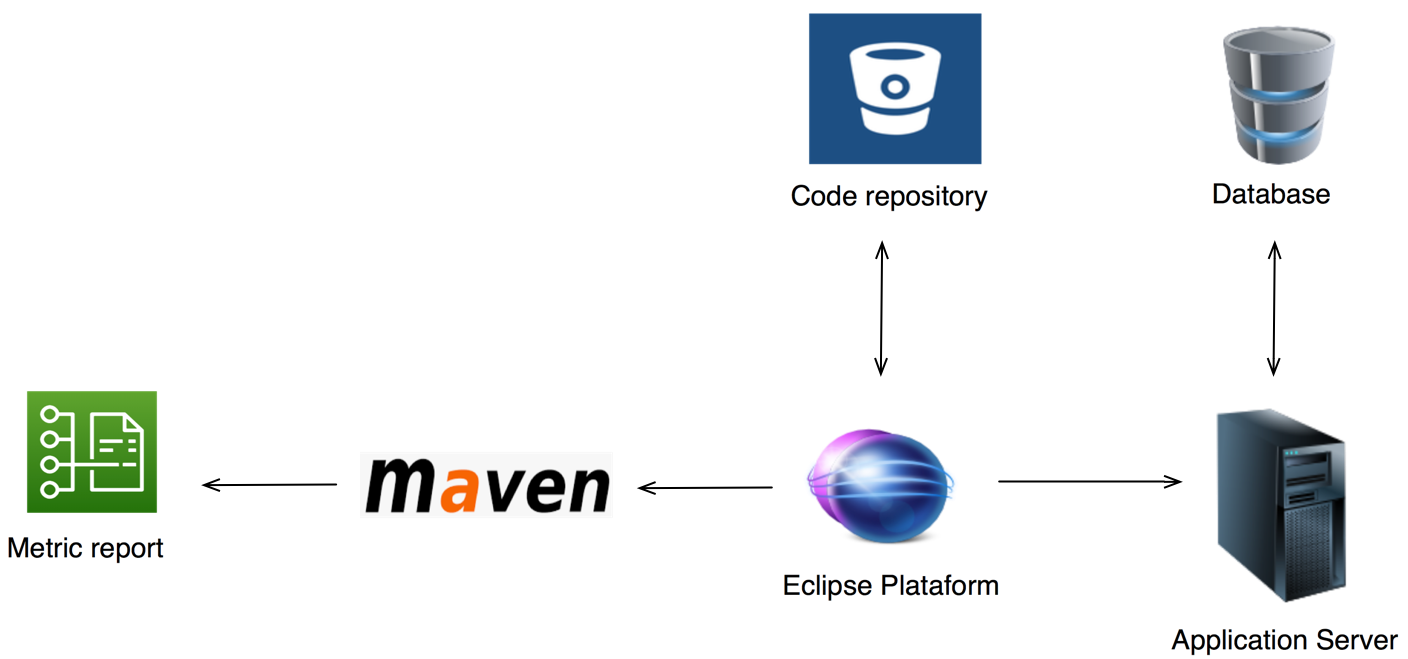}
    \caption{Development environment structure and metrics extraction}
    \label{figure:estrutura_do_ambiente_de_desenvolvimento_e_extracao_de_metricas}
\end{figure}

For the development of the SPL, the Integrated Development Environment (IDE) MyEclipse was used, based on the Eclipse IDE, but with a commercial license, using the Mercurial versioning tool through the BitBucket versioning repository. The execution environment used consists of a Glassfish application server and a PostgreSQL database. Figure \ref{figure:estrutura_do_ambiente_de_desenvolvimento_e_extracao_de_metricas} shows the structure of the development environment and metrics extraction.

\section{Evaluation}
\label{section:avaliacao}

This section discusses the methodology followed to evaluate the proposed WebSPL. The evaluation seeks to compare products derived from WebSPL with web applications developed without the use of a product line. This comparison seeks to reveal the benefits of using SPL from the perspective of some metrics. For this, Section~\ref{subsection-LPs-desenvolvidas} discusses the SPL developed. Section~\ref{section:metricas} describes the metrics used in the evaluation. Section~\ref{section:discussao} discusses the results obtained.

\subsection{WebSPL developed}
\label{subsection-LPs-desenvolvidas}

To assess the benefits of using WebSPL, two web applications with the same functionalities were developed, which were evaluated using software metrics. The first application, named in this article as complete-web-application, was developed with all features previously mentioned (Table~\ref{table:versos_da_linha_de_produto_de_software}). The second application, named in this article as \textit{application-web-spl-full}, was fully developed using the proposed WebSPL, extracting a product derived from WebSPL. It is important to emphasize that the applications had the same architectural pattern, used the same technologies, and had the same functionalities.

\subsection{Used Metrics}
\label{section:metricas}

To collect the application metrics, the SonarQube\footnote{SobarQube: https://www.sonarqube.org/} tool was chosen, which is a robust, open source web application that provides a set of software metrics, recognized and used in the market, for example, complexity per class, number of lines per class, and others. Sonar allows: (1) metrics to be combined or configured; (2) new metrics can be added via plug-ins; and (3) allows metrics to be compared based on a historical basis.

No SonarQube metrics have been customized. Thus, the metric extraction used the standard Sonar Java language profile, which already has a set of predefined metrics. Sonar has a considerable number of metrics, which are grouped by category, including complexity, size, duplicity (or clone), and technical debt. According to the characteristics of the proposed SPL project, the following metrics were chosen:

\begin{itemize}
\item \textbf{Complexity per class}: It is the average cyclomatic complexity per class. Cyclomatic complexity is also known as the MacCabe metric and represents the number of different paths a method or class can take.
    
    \item \textbf{Lines of code}: Number of lines of code present in each project, excluding comments, blank lines, or lines of documentation. Through this number, it is possible to visualize the size of each project.
    
    \item \textbf{Package cycles}: Number of unwanted package cycles. This metric can be applied to projects, modules, or directories and indicates that there may be an unwanted number of dependencies indicating code coupling.
    
    \item \textbf{Duplicate lines}: Number of lines considered to be duplicated. Metric that can be used to indicate the need for code refactoring or finding code points that originated from snippets that were copied and pasted.
    
    \item \textbf{Technical debt}: Technical debt is a metric based on the open-source SQALE methodology, which can be summarized as the organization of non-functional requirements related to code quality. Through a code analysis looking for violations of rules defined in Sonar's quality profile and prioritizing in order of importance reusability, portability, maintainability, security, efficiency, readability, and testability, the technical debt metric is generated. The technical debt represents the effort based on the number of days that must be used to remove all the technical debts found, in this case, the default configuration used was that each day has eight hours of work.
\end{itemize}    
    
It is noteworthy that both compared projects have the same object-relational mapping (ORM), this approach, in addition to reusing the classes that represent the domain entities, avoids the distortion of the metrics in favor of any project. The sum of the metrics of the Web application that derives from SPL is composed of the sum of the metrics of the web project itself and the sum of the metrics of the features implemented in SPL. In practice, the derived web project did not influence the final sum of the metrics, as it has only one line of Java code, all features were added to the project using Maven for dependency management.

\subsection{Results obtained}
\label{section:discussao}

Table~\ref{table:metricas} presents the results obtained from the metrics used. It is possible to observe the increasing complexity, number of lines of code, as well as technical debt of SPL applications. On the other hand, there is a reduction in the number of lines of code, as well as a reduction in the occurrence of code duplication in SPL applications. It is also important to highlight that the formation of cyclical dependency between the packages that constituted the SPL-derived applications and the non-SPL-derived applications was not observed.

\textbf{Increase the complexity of applications with SPL.} The construction of an SPL through a hierarchical model increases the complexity of components due to the need to implement features in the most flexible way possible. This is necessary to propagate the use of features at various hierarchical levels that SPL can take over time~\cite{Dincel2006}. The increase in complexity can be seen in Table~\ref{table:metricas}, the size of complexity increased by 56 units in the case of SPL presented in this article. According to \cite{Gurp2010}, an SPL with many variations can make software development more complex and difficult to maintain.

\begin{table}[!ht]
    \footnotesize
    \caption{Collected results}
    \label{table:metricas}
    \begin{tabular}{l l c c c c}
        \toprule
            Category & Metric & CWA & SPL & DWA & SAWS \\
        \midrule
Complexity & Complexity per class & 447 & 503 & 0 & 503 \\
Size & Number of Code Lines & 3091 & 3324 & 1 & 3325 \\
Design & Package Cycles & 0 & 0 & 0 & 0 \\
Duplicity & Duplicate Lines & 186 & 100 & 0 & 100 \\
Technical Debt & Technical Debt Level & 10.6 & 12.2 & 0 & 12.2 \\
        \bottomrule
    \end{tabular}
    \begin{tabular}{l l l l l l }
        \textit{Legend:}                                           & & & & & \\
        AWC: Complete Web application                               & & & & & \\
        SPL: Software product line                                  & & & & & \\
        DWA: Derived Web application from the complete SPL              & & & & & \\
        SAWS: SPL and application-Web-SPL-full                          & & & & & \\
    \end{tabular}
\end{table}

\textbf{Increase in the number of lines of code.} The increase in complexity is directly associated with the lines of code metric, according to \cite{Pohl2005}. This is due to the need to satisfy customers\rq wishes. This way, adding more functionality in the code to satisfy the needs, with that more lines of code will be used. As highlighted by \cite{Capilla2003} \cite{Yoshida2006}, SPL leads to a reduction of lines of code in the final product, however, the authors do not account for SPL overhead on the product, focusing only on the evaluation of the final derived product. In the example presented in this article, there is a drastic reduction in the number of lines and other metrics extracted from the derived final product.

\textbf{Cycles between packages were not detected.} The design in an SPL architecture seeks to provide and increase the reuse of features, focusing on evolution and ease of maintenance, allowing products to be customized. Coupling is one of the metrics used to evaluate the design of the classes that make up the project, according to \cite{VanderHoek2003} coupling allows a view of how modular the software is, that is, how much the relationship between classes and packages is. they are independent. It is possible to observe that the coupling between packages remained the same in the derived web application and in the web application that does not use SPL, in both cases the coupling between packages reported by Sonar was equal to zero.

\textbf{Reduction in the number of duplicated code.} Code duplication should be avoided as much as possible, however, through Sonar, it was possible to observe that the lines considered duplicated in the SPL correspond to the model classes, more specifically in the hashCode by a method using an attribute with the same name. As it is highly recommended by the Java language that each class must have its hashCode method implemented separately, it is clear in this case that, despite being considered duplicity of code, it will not be necessary to perform the refactoring.

\textbf{Increase in technical debt.} Technical debt is due to the rush to deliver the software. Consequently, best practices are not used, which may lead to future rework. According to~\cite{SEI2014}, technical debt is a metaphor that conceptualizes the balance between short-term value and long-term value, decisions made in the short term influence the long term. The management of technical debt is an increasingly critical aspect, as through it it is possible to concretely discuss the value and priority of product quality. According to \cite{Kruchten2012} the SQALE analysis method\footnote{Software Quality Assessment based on Lifecycle Expectations} can be used as a tool to manage technical debt, as it has an approach based on code analysis using defined quality indicators by the International Organization for Standardization (ISO) and quality standards such as testability, maintainability, portability, etc. The technical debt reported by Sonar points a way for SPL evolution and improvement, helping not only to improve quality but quantifying in days the time of rework effort, thus supporting schedule adjustments and time estimates.

\textbf{Improved the maintainability of the proposed SPL.} Each feature was implemented and modularized in a single project. This was possible when using CDI and Maven to control the dependency injection process and systematically manage dependencies between features. This better modularization of features also provided better support for possible changes in SPL, typically in evolution and maintenance scenarios, something that was not found in applications not derived from SPL. When such changes happen, each feature can evolve separately, minimizing, as much as possible, the chances of unwanted propagation. This lesser tendency to propagate changes is observed when trying to change the code of a feature. All modifications end up being confined in the context of the feature code itself, they do not propagate to the code related to other features.

Finally, although there are different ways to implement an SPL, no approach similar to the one presented in this article was found in the literature. It is also observed that, with an approach using the standard Java API, it is not necessary to use a solution as an aspect-oriented approach, which is typically pointed out as a trend in SPL implementation.

\section{Conclusions and Future Work}
\label{section:conclusao}

This article presented a product line of software for web applications. The SPL presented in this article uses Maven to manage dependencies between features, allowing an accurate and automatic mapping of mandatory features present in the feature diagram, thus becoming available to the end product through the use of CDI dependency injection. The interaction between features enabled in SPL makes use of features present in the CDI API such as event support. As such, the coupling can be significantly reduced. Avoiding the use of aspect-oriented programming it is possible to implement an SPL that uses purely Java language syntax, without relying on frameworks external to the standard API.

With the use of software metrics, it was possible to measure the features and classes that makeup two web applications, comparing an ad hoc web application and an SPL-derived web application. It is clear that SPL requires greater care due to its addition of lines and, consequently, complexity and technical debt, but it is possible to observe a real gain in the resulting final product when compared to the ad hoc web application. According to \cite{Dincel2006}, the use of metrics is common in several areas and aims to help in different situations, learning from the past, evaluating the present, and, in some cases, predicting the future.

In the future, new features or new metrics can be added to enrich this study with a broader view. We can observe that this article is an initial step in the search for new ways of implementing SPL and in the search for continuous improvement, serving as a reference for evaluating new empirical studies, which can be compared in other real-world contexts. Finally, we hope that the questions presented throughout this article will encourage other researchers to carry out their studies following the quality model proposed here and also evaluate it in the future, in different contexts.

\bibliographystyle{ACM-Reference-Format}
\bibliography{sample-base}

\end{document}